# Fault Diagnosis for Distributed Systems using Accuracy Technique

Poorva Kulkarni[1], Varsha Deshpande[1], Latika Sarna[1], Sumedha Shenolikar[1], and Supriya Kelkar[1]

[1] Computer Department, Cummins College of Engineering for Women, Karvenagar, Pune-411052, India.
{poorva.kulkarni, varsha.deshpande1, latika.sarna, sumedha.shenolikar, supriya.kelkar}@cumminscollege.in

**Abstract.** Distributed systems involve two or more computer systems which may be situated at geographically distinct locations and are connected by a communication network. Due to certain failures in the communication link, or the system itself, faults arise which may make the entire system dysfunctional. To enable seamless operation of the distributed system, these faults need to be detected and located accurately. This paper examines various techniques of handling faults in distributed systems and proposes an innovative technique which uses of percent accuracy for detecting faulty nodes in the system. Every node in the system acts as an initiator and votes for certifying faulty nodes in the system. This certification is done on the basis of percent accuracy value of each faulty node which should exceed a predefined threshold value to qualify the node as faulty. As the threshold value increases, the number of faulty nodes detected in the system reduces. This is a decentralized approach with no dependency on a single node to act as a leader for diagnosis. This technique is also applicable to ad-hoc networks, which are static in nature.

## 1 Introduction

Distributed systems are composed of many different individual machines which are connected to each other in a network. Due to the various events and dynamics of the systems, many faults may occur. These faults can either be connection faults or individual computer system faults. Distributed Systems performing real-time operations may face catastrophic situations if faults are not detected in time. This may affect the time critical operations thus degrading the quality of service of the system. To ensure efficient and reliable working of the distributed system, fault diagnosis needs to be performed accurately. There has been ample amount of research in fault diagnosis of distributed systems. This paper proposes a fault diagnosis algorithm which uses 'Accuracy' technique. The algorithm provides as an output, the percent value of particular node being declared as faulty. This value is used to estimate accurately which node is most likely to be faulty by using the votes of its neighboring nodes. Further in this paper, Section 2 explores related work being done in this domain, Section 3 describes the Fault Diagnosis of Distributed Systems using 'Accuracy' Technique (FDD-A) in detail demonstrating the working of the same using an example. Section 4 shows the implementation details of FDD-A and conclusion is given in Section 5.

## 2   Related Work

There are various methods and techniques available in the earlier work for diagnosing the faulty nodes in distributed networks. Some of these methods are summarized and examined below.

Work done by Zhao, Liu, Liu, He, Wang proposes a graph-based fault diagnosis technique based on graph theory in distributed systems [1]. This algorithm represents a multi-relational graph. It is different from the traditional approach of testing system from outside in case of some availability failure. The proposed system uses internal monitoring of availability status information and provides a global fault diagnosis based on graph. This method mines faults in simulated and real datasets [1].

A comparison-based system level fault diagnosis by Mourad and Nayak uses a Neural Network approach [2]. In a situation where all comparison outcomes are not available at the beginning of the diagnosis, it has been observed that a set of faulty nodes is not identified efficiently. Neural Network based self-diagnosis approach takes partial syndrome into consideration thus providing better and efficient results [2].

Leslie Lamport proposes a general approach that implements distributed systems having any suitable level of tolerance of faults [3]. In this technique, a clock-driven algorithm is executed by the processes which is simpler compared to explicit timeouts. The approach assumes a clock synchronization which is reliable and a key to the Byzantine general problem. This approach handles clock driven issues of the commands. A node in the network need not wait for acknowledgement from other nodes and can operate using clocks in parallel. In contrast to the algorithms seen before, Ferdowsi, Jagannathan, and Zawodniok have presented a new approach to fault detection using Outlier analysis [4]. Outlier analysis is the process of identifying anomalies or inconsistencies in the data, which can be a contemporary solution to fault diagnosis. In order to evaluate the actual system state and simultaneously to identify and remove the faults, a feed-forward neural networks (NN) has been considered. This system has been proven to identify and remove faults with a steadily high performance [4].

A distinct distributed fault diagnosis is suggested by Duarte, Weber and Fonseca [5]. In this work, Distributed Network Reachability (DNR) algorithm has been proposed. DNR demonstrates the nodes which are reachable and unreachable. This factor is essential to avoid sending of messages to unreachable nodes in distributed systems. The topology is arbitrary and dynamic, and the faults considered are crash faults as well as timing faults. The links of the network are checked by either of the connected nodes; alternately after certain intervals of time [5].

Punyotoya and Khilar have proposed a fault diagnosis algorithm for dispersed clusters [6]. The network is considered to have an arbitrary topology with k-connectivity, where k designates the number of clusters. There are a series of intermediate nodes which send messages between the desired source and destination. An online, two-phase algorithm is explained which gives a fault model based on heartbeat signals. The algorithm is dynamic and handles various situations of a distributed network, with diagnostic latency of O (1) thus, making it efficient [6]. Djelloul, Sari and Sidibe proposed a technique for fault diagnosis in manufacturing systems [7]. This technique ensures maintenance and proper functioning of the systems. There are three sections in the proposed system namely detection, diagnosis and decision-making sections. Data mining techniques used here help in repair activity in the decision-making process. For accurate detection of faults, classification methods based on hybrid neural networks are used [7].

Sreerama and Swarup have proposed an innovative approach of fault detection using Petrinets [8]. It mainly focuses on fault detection in a very short time interval. As many fault detection algorithms used network topology-based approach, the real fault distance and the calculated fault distance were found to be different. Petrinets prove to be a solution to this problem as they do not use iterations. They can be used for distributed networks and also for single fault and multiple faults detection [8].

Lala, Karmakar and Ganguly present a fault detection algorithm for distributed systems based on localization bounded by time and frequency [9]. This approach uses Stationary Wavelet Transforms and includes a feature extraction process from the signals. They have also used machine learning and artificial neural network to detect the faults. This algorithm proves to be accurate in fault detection and localization [9]. Benayas, Carrera and Iglesias have presented an approach based on Software Defined Networks (SDN) [10]. Nowadays many networking problems are solved using SDN. Though the approach is more dynamic in nature, it also introduces potential faults in the system and hence it becomes necessary to rectify them. In this paper a Machine learning approach has been introduced. Also, Data Analysis is used to monitor the SDN networks. This automates the fault detection task. It aims to develop humans with high skills to operate the system and detect the faults automatically [10].

In the domain of distributed systems, fault prevention can be more effective than fault diagnosis. In the work presented by Waszecki, Kauer, Lukasiewycz, Chakraborty, four distinct early fault detection methods have been proposed [11]. If the intermittent faults are taken care of in advance, precautionary measures can be taken to avoid permanent component failures [11]. Tran and Schönwälder have proposed DisCaRia which is a reasoning system [12]. DisCaRia helps in rectifying faults by using knowledge resources of faults. This system can be implemented using peer to peer technology [12].

Bastida and Chukhrey have presented a fault diagnosis algorithm for a gyroscopic sensor unit [13]. In this algorithm, Fault Location takes place after continuous monitoring of the system. Once the faulty sensor is located, the next step involves identifying the type of the fault in the sensor [13]. In the work introduced by Cui, Shi and Wang, a fault diagnosis algorithm for cyclic and temporal networks is suggested [14]. This approach provides the precise information about fault causes and gives time intervals of fault occurrences. It helps in identifying the fault causes and the components vulnerable to the faults [14]. Horii, Kobayashi, Matsushima and

Hirasawa have proposed Lagrangian Relaxation based Algorithm for fault diagnosis [15]. This algorithm works on Linear Programming (LP) and is applied to multicomputer systems. It is faster and accurate in error detection [15].

Earlier work proposed a new technique where two nodes acting as coordinators diagnose the system for faulty nodes [16]. Although this algorithm is static in nature, it ensures that a 'single point' failure does not affect the diagnosability of the system. The algorithm presented by Kelkar and Raj Kamal suggests detecting faulty nodes on Controller Area Network [17]. This approach proves that it uses a well-defined and bounded number of testing rounds and messages for the completion of one diagnostic cycle. The Leader based fault diagnosis algorithm proposed by Manghwani, Taware, Kelkar, Chinde and Alwani includes selection of a leader in the diagnostic cycle [18]. All the nodes other than the leader node send their diagnostic information to the leader node. Based on whether the information packet is received or not, the corresponding node is detected as faulty or faulty-free and this information is maintained and recorded with the leader node.

The algorithm proposed by the authors of this paper is an Accuracy-based algorithm. This uses neighbour recommendation to help certify the node as faulty. The voting technique and threshold value ensure that nodes are correctly classified as faulty.

## 3   Proposed Fault Diagnosis Algorithm for Distributed Systems using Accuracy Technique

The system considered for fault diagnosis is not fully connected, which means all the nodes in the system are not connected to each other. The given system is static. Static networks are those where the topology and connection between the fixed number of nodes is already known to all the nodes in the system. In this algorithm, every node knows the entire system.

The proposed system consists of seven nodes namely - $N_1$, $N_2$, $N_3$, $N_4$, $N_5$, $N_6$ and $N_7$. These nodes are connected in an arbitrary topology as shown in Fig. 1.

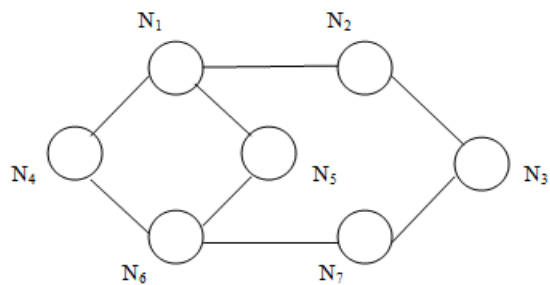

**Fig. 1.** System considered for fault diagnosis.

| IP Address | Status |
|---|---|

**Fig. 2.** Status Frame

| Node i | Vote i | Faulty Node IP Address i | Voter Node IP Address i | Reachability of faulty node i |
|---|---|---|---|---|
| Node j | Vote j | Faulty Node IP Address j | Voter Node IP Address j | Reachability of faulty node j |
| | . | . | . | . |
| | . | . | . | . |
| | . | . | . | . |
| Node n | Vote n | Faulty Node IP Address n | Voter Node IP Address n | Reachability of faulty node n |

**Fig. 3.** Fault Count Frame.

**Table 1.** Reachability Table.

| Nodes | Reachability |
|---|---|
| $N_1$ | 3 |
| $N_2$ | 2 |
| $N_3$ | 2 |
| $N_4$ | 2 |
| $N_5$ | 2 |
| $N_6$ | 3 |
| $N_7$ | 2 |

The FDD-A algorithm operates in three stages namely, Status Acquisition, Fault Count Frame Transfer and Fault Qualification. The fault certification takes place on the basis of accuracy. Accuracy can be defined as the measure of preciseness with which the neighbour nodes certify a particular node as faulty. Reachability of any given node is defined here as the maximum number of direct one to one connections it has with the other nodes of the network. The initiator is determined based on the priority of the node found first in the reachability table (Table 1.) which is stored at each node, having highest reachability. Accuracy can be calculated as the ratio of number of votes a particular node obtains certifying it as faulty, to the reachability of the node.

## 3.1 Details of Fault Diagnosis for Distributed System using Accuracy Technique (FDD-A)

### 3.1.1 Status Acquisition

*3.1.1.1 Self-Test*

Each node will self-test itself and if it is fault-free it will exchange hello messages with its neighbours. Self-test is a series of arithmetic, logical and memory operations which test the correctness of the processor of every node.

*3.1.1.2 Reception of neighbours' status frames*

The hello message requests the neighbours to send their fault status to the requesting node. The nodes which have correct self-test results send '0' indicating their fault-free standing and those nodes which fail the self-test send '1' denoting their faulty status. This status information is sent to the requesting node in the form of a status frame which is shown in Fig. 2. There may be situations where faulty nodes are unable to send their fault status due to being faulty.

*3.1.1.3 Updation of Local Status Frame*

After receiving status frames from neighbour nodes, the requesting node will update its local status frame based on the status frames sent by neighbour nodes to it. The requesting node then writes the status and IP addresses of all its neighbours into its local status frame. In some cases, requesting node may not get the status frames from the faulty neighbor nodes within the prescribed time. The requesting node will mark the status of such non-responding nodes as faulty nodes in its local status frame. This status frame is further used in rendering the creation or updation of the Fault Count Frame as shown in Fig. 3, based on the faulty nodes. Thus, stage 1 is performed by every node in the system.

### 3.1.2 Fault Count Frame (FCF) Transfer

*3.1.2.1 Creation or Updation of FCF at the initiator node*

The fault-free node having the maximum reachability count starts the diagnosis cycle. Let us call this node as the initiator. Initiator will begin by updating the Fault Count Frame (FCF) based on the status frame received from its neighbour nodes. As shown in Fig. 3, there are four fields in the FCF namely, vote, faulty node IP address, voter node IP address and reachability of the faulty node. Initiator will append FCF only when it finds a faulty neighbour. If any neighbour node is found to be faulty, then initiator will now act as a voter by incrementing the vote field and adding IP address of faulty node along with its own IP address as voter node address and finally adding the reachability count of the faulty node.

*3.1.2.2 Selection of next initiator*

Now the initiator will select the next initiator by iterating the reachability table i.e. Table I by selecting the node with the next highest reachability or the node with same reachability count but, with lower priority. Once the next initiator has been selected, the current initiator will pass the FCF to it. The next initiator will continue the diagnosis by checking the status frames received from its neighbour nodes. It appends the FCF as discussed in Step 1 of the FCF Transfer stage. If the current initiator finds the faulty node entry which is already available in the FCF, it only increments the vote count without adding a duplicate entry.

*3.1.2.3 Circulation of FCF to all nodes*

Step 2 of the FCF transfer stage is repeated till FCF is passed to all the nodes of the system thus making every node of the system as initiator.

### 3.1.3 Fault Qualification

*3.1.3.1 Calculation of percent accuracy*

At the end of step 3 of the FCF transfer stage, the most recent initiator calculates the percent accuracy for each faulty node found in FCF.

$$\text{Percent Accuracy of a faulty node} - (\text{Votes} / \text{Reachability}) * 100 , \qquad (1)$$

*3.1.3.2 Comparison of faulty node percent accuracy with percent threshold value*

The percent accuracy of a faulty node is compared with percent threshold value, which is currently predefined for the algorithm as 75%. This predefined value of percent threshold can be decided based on the type and the severity of the faults in the system and their effects on the operation of the system. If percent accuracy of a faulty node is greater than or equal to the percent threshold value, then it is certified as faulty, else as fault-free node. This certification is done by the most recent initiator node.

*3.1.3.3 Broadcasting faulty nodes information to all the nodes*

After certifying the faulty nodes, the most recent initiator broadcasts the list of certified faulty nodes to all other nodes in the system. This will facilitate the faulty nodes information to be spread throughout the network. Once the nodes are certified as faulty, they will be barred from performing any operations until they are repaired.

## 4  Implementation Details

The system considered for implementation consists of seven nodes as shown in Fig. 1. The Table 2. shows the mapping of the nodes in the Fig. 1 to the corresponding IP Addresses used for implementation.

Fig. 4 shows the formation of status frame by initiator $N_6$. As shown in Fig. 1, $N_6$ has three neighbours, namely $N_4$, $N_5$ and $N_7$. Thus, the status frame indicates that the node $N_7$ is faulty and sets its status information as '1'.

Fig. 5 shows screenshot of initiator $N_6$ while preparing FCF. As seen in Fig. 5 part (1), initiator $N_6$ receives status from its three neighbours $N_4$, $N_5$ and $N_7$. Depending upon status frames received, $N_6$ appends the FCF as shown in Fig. 5 part (2). $N_7$ is found to be faulty and its information is added in FCF.

**Table 2.** IP Addresses of the nodes in the network.

| Nodes | IP Addresses |
|---|---|
| $N_1$ | 172.16.27.103 |
| $N_2$ | 172.16.27.104 |
| $N_3$ | 172.16.27.105 |
| $N_4$ | 172.16.27.106 |
| $N_5$ | 172.16.27.107 |
| $N_6$ | 172.16.27.108 |
| $N_7$ | 172.16.27.110 |

```
172.16.30.106,0
172.16.30.107,0
172.16.30.110,1
```

**Fig. 4.** Status Frame at Node $N_6$.

```
ccoew@client1: ~/Desktop/Accuracy
ip 172.16.30.108
faultfree
ip of me <type 'str'>
ipofmax 172.16.30.108
maximum count is 3
I have the max reachability count hence, prepare fcf
['172.16.30.106', '172.16.30.107', '172.16.30.110', '']
My status is: 0
Receiving neighbour's status
Received status from neighbour 0
Receiving neighbour's status
Received status from neighbour 0
Receiving neighbour's status
Received status from neighbour 1
Line1 ['172.16.30.106,0', '172.16.30.107,0', '172.16.30.110,1', '']
array eth1
ip 172.16.30.108
faultfree
My IP address:  <type 'str'>

One faulty node found...putting into fcf
FCF Prepared for the first time by current node ===>  [{'vote': 1, 'ip': '172.1
6.30.110', 'from_ip': '172.16.30.108', 'rch': '2'}]
```

**Fig. 5.** Formation of the Fault Count Frame.

```
FCF received from previous node: [{'vote': 2, 'ip': '172.16.30.110', 'from_ip':
'172.16.30.108', 'rch': '2'}, {'vote': 2, 'ip': '172.16.30.104', 'from_ip': u'17
2.16.30.103', 'rch': '2'}]
One faulty node found which was not present before. Hence, putting into fcf
reach_faulty 3
ip_faulty 172.16.30.108
```

**Fig. 6.** Reception of FCF from previous initiator

```
FCF UPDATED BY CURRENT NODE :  [{'vote': 2, 'ip': '172.16.30.110', 'from_ip': '1
72.16.30.108', 'rch': '2'}, {'vote': 2, 'ip': '172.16.30.104', 'from_ip': u'172.
16.30.103', 'rch': '2'}, {'vote': 1, 'ip': '172.16.30.108', 'from_ip': '172.16.3
0.107', 'rch': '3'}]
```

**Fig. 7.** Updation of FCF.

Fig. 6 shows the reception of FCF by the node $N_5$. Here, the FCF has two nodes present in it, which were detected as faulty by the previous initiators when the FCF was passed to them. The 'from_ip' field indicates the IP address of the node which identified the corresponding node in the FCF as faulty. Node $N_5$ further checks its neighbours' status and diagnoses $N_6$ as faulty. On checking the FCF, it realizes that $N_6$ is yet to be called as faulty i.e. it is not present in the FCF. Hence, $N_5$ appends status information of $N_6$ in the FCF. Had $N_6$ been previously present in FCF, $N_5$ would have just incremented the vote count of $N_6$.

Fig. 7 demonstrates the further concatenation of FCF by node $N_5$. $N_5$ appends the faulty node $N_6$ with IP address of 172.16.30.108 to the FCF. This shows the previously appended nodes which were diagnosed as faulty, as well as information updated by $N_5$ about the fault status of $N_6$. The updated FCF with three faulty nodes will be passed on to the next highest reachability node which will act as the next initiator.

```
Final IP for calculation:   172.16.30.110
Votes of IP for calculation:   2.0
Reachability of faulty node for calculation:   2.0
Percentage:   100.0

Final IP for calculation:   172.16.30.104
Votes of IP for calculation:   2.0
Reachability of faulty node for calculation:   2.0
Percentage:   100.0

Final IP for calculation:   172.16.30.108
Votes of IP for calculation:   1.0
Reachability of faulty node for calculation:   3.0
Percentage:   33.3333333333

=========== NODES QUALIFIED AS FAULTY ARE ==========
 ['172.16.30.110', '172.16.30.104']
Sent Broadcast of faulty nodes to other nodes.
ccoew@client1:~/Desktop/Accuracy$
```

**Fig. 8.** Final Classification of faulty nodes.

```
BROADCAST OF FAULTY NODES IN NETWORK RECIEVED ===> ['172.16.30.110', '172.16.30.
104']
ccoew@server1:~/Desktop/Accuracy$
```

**Fig. 9.** Broadcast received by other nodes.

Fig. 8 shows the screenshot captured at node $N_5$. The calculation for percent accuracy and certification of faulty node/s is shown. The screenshot shows the calculation of nodes $N_7$, $N_2$ and $N_6$ respectively performed at the most recent initiator node $N_5$. Their percent accuracy is compared with the percent threshold value and the certification of faulty node/s is done accordingly by node $N_5$.

As seen in the screenshot, nodes $N_7$ and $N_2$ have percent accuracy as 100% as all of their neighbours have correctly voted them as faulty. Thus, when the percent accuracy of $N_7$ and $N_2$ are compared with percent threshold value which is 75%, they are certified as faulty by the node $N_5$, as their percent accuracy is greater than their percent threshold value. But the percent accuracy of $N_6$ is 33.33% which is less than percent threshold value. Therefore, $N_6$ is not certified as faulty by the node $N_5$. The certified faulty nodes are then broadcasted by the most recent initiator $N_5$ to all other nodes in the system.

Fig. 9 indicates the broadcast received by other nodes sent by the most recent initiator. Hence, each node, whether faulty or fault-free will receive complete network information, to assist in taking decisions regarding load redistribution.

Table 3. shows the results of a probable implementation. The value for number of votes is considered for the purpose of analysis. Based on these assumptions about the votes, the percent accuracy is calculated using equation (1). Referring to this table a graph is plotted for different threshold values.

Fig. 10 depicts a graph of number of nodes classified as faulty against the percent threshold. Percent threshold value is nothing but maximum percent accuracy value within which a node is treated as fault-free. Based on Table III, the number of faulty nodes varies with respect to the change in percent threshold value. It is observed that the number of nodes to be certified as faulty decreases as the percent threshold increases.

**Table 3.** Votes per node and percent accuracy.

| Nodes | Votes/Reachability | % Accuracy |
|---|---|---|
| $N_1$ | 1/3 | 33.33 |
| $N_2$ | 1/2 | 50 |
| $N_3$ | 1/2 | 50 |
| $N_4$ | 1/2 | 50 |
| $N_5$ | 0/2 | 0 |
| $N_6$ | 2/3 | 66.67 |
| $N_7$ | 2/2 | 100 |

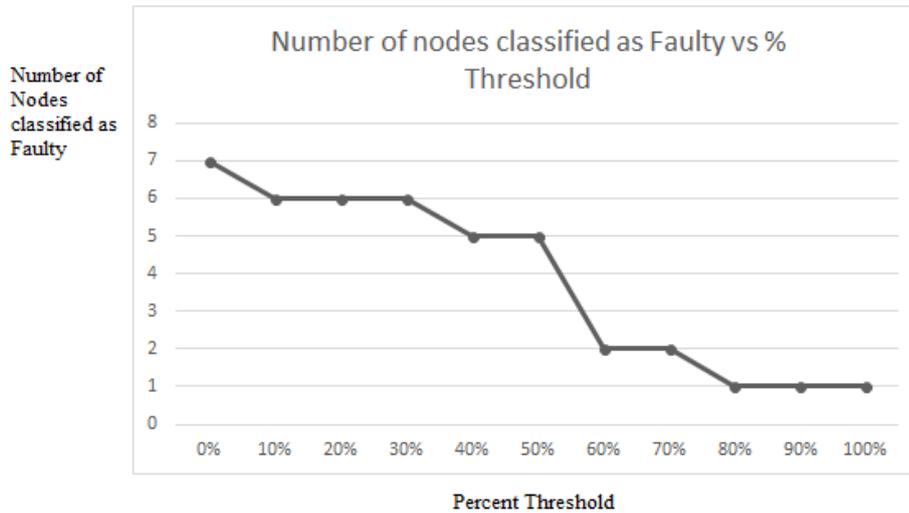

**Fig. 10.** Number of nodes classified as faulty vs Percent Threshold Value.

## 5   Conclusion

The algorithm proposed devises a new method of fault diagnosis which is an innovative approach as compared to its counterparts. Instead of relying only on the self-test results, this algorithm also takes into consideration the votes of the neighbour nodes, while declaring a particular node as faulty. This guarantees the accurate detection process of the faulty nodes every cycle. The information thus rendered can further be used while implementing load balancing, availability and reliability utilities or roles of a distributed network. Although static in nature, this algorithm can work on any arbitrary topology once the reachability of each node is known by every other node. Thus, fault diagnosis is done precisely, with the recommendation of neighbour nodes above a certain fault threshold. The proposed algorithm does not handle communication link failures. Future work includes detecting faults using this approach in dynamic networks.

## References

1. Zhao, L., Liu, Z., Liu, W., He, H., Wang, Y.: G-FDDS: A Graph-based Fault Diagnosis in Distributed   Systems, 2nd IEEE International Conference on Computational Intelligence and Applications (2017) 559–567
2. Elhadef, M., Nayak, A.: Comparison-Based System-Level Fault Diagnosis: A Neural Network Approach, IEEE Transactions on Parallel and Distributed Systems, Vol. 23, Issue 6 (2012) 1047–1059

3. Lamport, L: Using Time Instead of Timeout for Fault-Tolerant Distributed Systems, ACM Transactions on Programming Languages and Systems, Vol. 6, Issue 2 (1984)
4. Ferdowsi, H., Jagannathan, S., Zawodniok, M.: An Online Outlier Identification and Removal Scheme for Improving Fault Detection Performance", IEEE Transactions on Neural Networks and Learning Systems, Vol. 25, No. 5 (2014) 908–919
5. Duarte, E., Weber, A., Fonseca, K.: "Distributed Diagnosis of Dynamic Events in Partitionable Arbitrary Topology Networks", IEEE Transactions on Parallel and Distributed Systems, Vol. 23, No 8 (2012) 1415–1426
6. Punyotoya, S. Khilar, P.: A Novel Fault Diagnosis Algorithm for K Connected Distributed Clusters, International Conference on Industrial Electronics, Control and Robotics December (2010) 101–105
7. Djelloul, I., Sari, Z., Sidibe, I.: Fault diagnosis of manufacturing systems using data mining techniques, 5th International Conference on Control, Decision and Information Technologies (CoDIT) (2018) 198 – 203
8. Sreerama, R., Swaru, S.: Detection, localization and fault diagnosis using PetriNets for smart power distribution grids, 7th International Conference on Power Systems (ICPS) (2017) 596 – 600
9. Lala, H., Karmakar, S., Ganguly, S.: Fault diagnosis in distribution power systems using stationary wavelet transform and artificial neural network, 7th International Conference on Power Systems (ICPS) (2017) 121–126
10. Benayas, F., Carrera, A., Iglesias, C.: Towards an autonomic Bayesian fault diagnosis service for SDN environments based on a big data infrastructure, Fifth International Conference on Software Defined Systems (SDS) (2018) 7-13
11. Waszecki, P., Kauer, M., Lukasiewycz, M., Chakraborty, S.: Implicit intermittent fault detection in distributed systems, 19th Asia and South Pacific Design Automation Conference (ASP-DAC) (2014) 646-651
12. Tran, H., Schönwälder, J.: DisCaRia—Distributed CaseBased Reasoning System for Fault Management, International Young Scientists Forum on Applied Physics and Engineering IEEE Transactions on Network and service management, Vol 12, No.4 (2015) 540-553
13. Bastida, J.P.M., Chukhray, A.G.,: An active Diagnostic Algorithm for a Gyroscopic Sensors Unit, International Young Scientists Forum on Applied Physics and Engineering (2016) 29-32
14. Cui, Y., Shi, J., Wang, Z., Fault Propagation Reasoning and Diagnosis for Computer Networks Using Cyclic Temporal Constraint Network Model, IEEE Transactions on Systems, Man and Cybernetics (2017) 1-14
15. Horii, S., Kobayashi, M., Matsushima, T., Hirasawa, S., Fault Diagnosis Algorithm in Multi-Computer Systems based on Lagrangian Relaxation Method, 2012 International Symposium on Information Theory and its Applications (2012) 712-716
16. Kelkar, S., Yeole, D.G., Sinkar, M.B., Jagtap, P.B., Zagade, D.S.: Coordinator- Based Adaptive Fault Diagnosis Algorithm for Distributed Computing Systems, International Conference on Advances in Computing, Communications and Informatics (ICACCI) (2017) 745-751
17. Kelkar, S., Kamal, R.: Adaptive Fault Diagnosis Algorithm for Controller Area Network, IEEE Transactions on Industrial Electronics, Vol. 61, No. 10 (2014)
18. Manghwani, J., Taware, R., Kelkar, S., Chinde, P., Alwani, S.: Leader Based Adaptive Fault Diagnosis Algorithm for Distributed Systems, 2017 International Conference on Information, Communication, Instrumentation and Control (ICICIC) (2017) 1-6